\documentstyle[twocolumn,aps,prl,epsf]{revtex}

\begin{document}

\title{{\bf Thermal Folding and Mechanical Unfolding Pathways of Protein
Secondary Structures }}

\author{{\bf Marek Cieplak$^{1,2}$, Trinh Xuan Hoang$^3$, and
Mark O. Robbins$^1$}}

\address{
$^1$ Department of Physics and Astronomy, The Johns Hopkins University,
Baltimore, MD 21218\\ 
$^2$ Institute of Physics, Polish Academy of Sciences,
Al. Lotnik\'ow 32/46, 02-668 Warsaw, Poland \\
$^3$ International School for Advanced Studies (SISSA),
via Beirut 2-4, 34014 Trieste, Italy}

\maketitle

\vskip 40pt
\noindent
$^*$Correspondence to: \\
Marek Cieplak,\\
Institute of Physics, \\
Polish Academy of Sciences, \\
Al. Lotnik\'ow 32-46  \\
02-668 Warsaw, Poland\\
Tel:  48-22-843-7001\\
Fax:  48-22-843-0926\\
E-mail: mc@ifpan.edu.pl

\vskip 40pt
\noindent
Grant sponsor: NSF DMR-0083286, TIPAC (Johns Hopkins), 
and KBN (Poland) -  2P03B-146-18.


\newpage
\begin{abstract}
{Mechanical stretching of secondary structures is studied
through molecular dynamics simulations of a 
Go-like model.
Force vs. displacement curves are studied as a function of the
stiffness and velocity of the pulling device.
The succession of stretching events, as measured
by the order in which contacts are ruptured, is compared to
the sequencing of events during thermal folding and unfolding.
Opposite
cross-correlations are found for an $\alpha$-helix
and a $\beta$-hairpin structure.
In a tandem of two $\alpha$-helices, the
two constituent helices 
unravel nearly simultaneously.
A simple condition for simultaneous vs. sequential unraveling of
repeat units is presented.
}
\end{abstract}



\section*{INTRODUCTION}

Weak noncovalent bonding forces govern functioning and structural 
cohesion in cells.
Direct measurements of these forces through mechanical means
has recently become an important tool in studies of biological 
molecules. There is a variety of techniques for probing forces
in the pico- and nano-Newton range \cite{Leckband} such as atomic force
microscopy\cite{Figdor,Hoh,Hoh1,Evans}, 
optical tweezers \cite{Bustamante,Simmons}, 
the surface force apparatus \cite{Israelachvili},
micropipette aspiration\cite{aspiration}, and
the quartz microbalance \cite{Krim,Kasemo}.
As examples of recent achievements, we may list elucidation
of the nature of interactions of a chaperone protein (HIV-1)
with DNA through stretching of a strand of the DNA
with optical tweezers
\cite{Bloomfield} and discovery of stick-slip motion when
two strands of a DNA double helix are pulled apart \cite{Heslot}. \\

The techniques used in mechanical unfolding of individual
biological molecules rely on tethering of the molecule between
movable surfaces. This tethering is relatively easy to accomplish with
long molecules such as DNA and giant proteins such as titin 
\cite{Bustamante,Simmons} which are naturally built as a tandem array
of many globular domains. For shorter molecules, the pulling surfaces
interact and affect the pulled molecule in a way that makes the data hard to
interpret. In order to extend the method to single domained proteins,
Yang et al. \cite{Busamante1} have recently developed a method
of synthesizing identical repeats of protein molecules in the solid state
which were then studied using a modified scanning force microscope.
This technique has been applied to T4 lysozyme. \\

At this moment, experimental data on the mechanical unfolding 
of the secondary structures of proteins are not available.
However, data on periodically repeated proteins and even individual proteins
may become available in the near future.  From a
theoretical point of view, it is important to
gain an understanding of the basic unfolding mechanisms of
simple structures and to develop analytical tools
that could then be used for large proteins.
This process is facilitated by considering simple models that allow
a rapid exploration of parameter space.
Our choice in this paper is to analyze Go-like models \cite{Goabe} which 
emphasize the importance of native conformations and treat non-native
interactions only schematically.
The Go-like models \cite{Goabe},
though coarse-grained, are fairly realistic \cite{Stakada} in their kinetic
properties and allow for a thorough characterization and comparison of
mechanical, equilibrium and folding properties in a straightforward
manner.
This kind of full characterization is difficult to achieve in all-atom
models with the Amber \cite{Amber} or CHARMM \cite{Charmm} force fields
which nevertheless are well suited to studies of mechanical stretching.\\

The idea that
mechanical unfolding experiments on proteins have the potential to provide
insights into the relevant folding pathways is what motivated
Bryant et al. \cite{Rokhsar} to carry out all-atom (CHARMM-based)
simulations of the C-terminal hairpin of protein G, the folding of 
which has been previously studied experimentally by Munoz et al. 
\cite{Munoz,Munoz1}. They have found that, under low pulling forces, 
breakdown of hydrogen bonds 
precedes dissociation of the
hydrophobic cluster. Their interpretation of this finding is that
thermal folding should proceed in the opposite order
to mechanical unfolding.
If so,
then the zippering folding mechanism \cite{Munoz1} would be
less favored than one in which a hydrophobic cluster is formed first.
This prediction remains to be tested.\\

Here, we explore properties of
Go models of proteins through molecular dynamics simulations.
We consider the variant in which contact interactions
are described by Lennard-Jones potentials. The simulations 
include a Langevin noise term which
both mimics presence of a solvent and controls the 
temperature, $T$. This paper focuses on an $\alpha$ - helix of 16 monomers,
denoted as H16; a $\beta$ - hairpin of 16 monomers, B16;
and a double repeat of the $\alpha$ - helix, H16-2.
The companion paper \cite{second} describes a 
similar analysis for titin. \\

We first study
mechanical unfolding
at nearly zero temperature.
This choice of $T$ minimizes fluctuations and rate dependence,
and most simply reveals the effects of the structure of the energy landscape.
The results should be equivalent\cite{Evans} to fast stretching at higher $T$,
and temperature
dependence will be considered in subsequent work.
The protein is stretched by a Hookean cantilever and the force
is plotted as a function of the cantilever displacement.
We also characterize the stretching process by studying the succession
of unfolding events, which are described by the cantilever
displacements at which specific contacts are broken.
Both the force-displacement curve and the
order of unfolding events depend on the
stiffness and velocity of the cantilever.
We next discuss studies of folding, where temperature plays an essential role.
The sequencing of folding events depends on $T$,
and smooth and simple pathways are only found near
an optimal temperature denoted by $T_{min}$ 
\cite{Socci,Henkel,Hoang,Hoang1,optimal}.
The sequencing of folding events near $T_{min}$ is contrasted with 
that of stretching events for different protein structures.
We find that both sequencings are governed
primarily by the contact order \cite{Unger,Plaxco,Ruczinski}, 
i.e. by the distance between two
amino acids along the sequence of the protein.
However, the cross-correlations between thermal and mechanical sequencings
are opposite for the two simple cases considered: H16 and B16.
Only in the latter case do folding and stretching occur in the opposite order,
as envisioned by, e.g., Bryant et al. \cite{Rokhsar}.
In general, the thermal and mechanical pathways can be very different.
\\

Another quantity that we study here is what we propose to call
an irreversibility length, $L_{ir}$. If one studies folding  from a fully
extended conformation, then one finds that the characteristic folding time
diverges as $T\rightarrow$0.
Thus a fully stretched protein will not fold 
back to the native state at low temperatures.
On the other hand, a protein that is
pulled only slightly
will return to its native shape on release. There must then be
a characteristic stretched length of the protein, $L_{ir}$,
which separates the two behaviors.
We demonstrate that $L_{ir}$ does indeed exist and find that
it is substantially
larger for B16 than for H16. Furthermore, the folding time for
lengths less than $L_{ir}$ is a complicated function of
the mechanical extension.\\

We also consider a tandem arrangement of two $\alpha$-helices
and find that the constituent helices unravel almost simultaneously
whereas in titin \cite{second} the unraveling is serial in nature.
Simple criteria for the two types of behavior are described.\\


\section*{MODEL AND METHOD}

The model we use is described in detail in references 
\cite{Hoang,Hoang1,optimal,biolo}.
For simplicity, we consider the variant where steric constraints
associated with dihedral and other angles are ignored.
Briefly, a protein is modeled by a chain of identical beads
which correspond to the locations of the C${\alpha}$ atoms.
The consecutive beads interact through the potential \cite{Clementi}
\begin{equation}
V^{BB} = \sum_{i=1}^{N-1} [ k_1 (r_{i,i+1} - d_0)^2 +
k_2 (r_{i,i+1}-d_0)^4],
\end{equation}
where $r_{i,i+1}=|{\bf r}_i - {\bf r}_{i+1}|$ is the distance between
two consecutive beads, $d_0 = 3.8$ \AA, $k_1=\epsilon /$\AA$^2$ 
$k_2=100\epsilon /$\AA$^2$, and $\epsilon$ is the characteristic
energy parameter corresponding to a native contact.
The anharmonic term in equation (1) prevents energy localization
in specific phonons and thus accelerates equilibration \cite{Clementi}.
\\

The interaction that governs the native contacts
(defined as those C$\alpha$ that are not immediate neighbors, but
are no further than 7.5\AA\ apart in the native structure)
is chosen to be of the Lennard-Jones type (see e.g. \cite{Clementi1})
\begin{equation}
V^{NAT} = \sum_{i<j}^{NAT}4\epsilon \left[ \left( \frac{\sigma_{ij}}{r_{ij}}
\right)^{12}-\left(\frac{\sigma_{ij}}{r_{ij}}\right)^6\right],
\end{equation}
The parameters $\sigma_{ij}$ are chosen so that each contact in the native
structure is stabilized at the minimum of the potential,
and $\sigma \equiv 5$\AA\ is a typical value.
As a technical criterion for determining when a native contact forms or breaks
during the time evolution, we adopted
the cutoff value of 1.5$\sigma _{ij}$. 
The non-native contacts are described by purely repulsive potentials.
These are obtained by evaluating $V^{NAT}$ with a length parameter
$\sigma$, truncating the potential at its minimum ($2^{1/6}\sigma$),
and shifting it to have zero value at this cutoff distance.
\\

Figure 1 illustrates the forms of the potentials for the
$\alpha$-helix. 
When studying the folding times, we have adopted a simplified approach
in which a protein is considered folded if all beads that form a native
contact are within the cutoff distance of $1.5 \sigma _{ij}$ instead of making
a more precise delineation of the native basin as in ref. \cite{Hoang1}.
This will allow for a more meaningful comparison with the results
on titin \cite{second}. \\

The beads are coupled to Langevin noise and damping terms to
mimic the effect of the surrounding solution and maintain constant
temperature $T$.
The equations of motion for each bead are
\begin{equation}
m\ddot{{\bf r}} = -\gamma \dot{{\bf r}} + F_c + \Gamma \ \ ,
\end{equation}
where $m$ is the mass of the amino acids represented by each bead,
$F_c$ is the net force due to the molecular potentials and external forces,
$\gamma$ is the damping constant, and
$\Gamma$ is a Gaussian noise term with dispersion
$\sqrt{2\gamma k_B T}$.
We measure time in units of
the characteristic period of undamped oscillations
in the Lennard-Jones potential
$\tau \equiv \sqrt{m \sigma^2 / \epsilon}$.
Using typical values for the average amino acid mass,
length and binding energy yields
3$ps$ as an estimate of $\tau$.
According to Veitshans et al. \cite{Veitshans},
realistic estimates of damping by the solution correspond to a
value of $\gamma$ near 50 $m/\tau$.
However, the folding times have been found to depend on
$\gamma$ in a simple linear fashion for $\gamma > m/\tau$
\cite{Hoang,Hoang1,thir}.
Thus in order to
accelerate the simulations, we work with $\gamma \;= \; 2 m/\tau$.
The equations of motion are
solved by means of the fifth order Gear predictor-corrector
algorithm \cite{Gear} with a time step of $0.005\tau$.
\\

In order to pull the protein apart,
we attach both of its ends to purely harmonic springs of spring constant $k$.
We focus on three cases:
a) the stiff spring: $k \ge 60 \epsilon / $\AA$ ^2$ b) the soft spring:
$k\;=\;0.12 \epsilon / $\AA$ ^2$, and 
c)the very soft spring: $k\;=\;0.04\epsilon / $\AA$ ^2$.
The outer end of one spring is held stationary, and the other is pulled
at a fixed rate $v_p$.
This models stretching by a Hookean cantilever with stiffness $k/2$,
since the two springs add in series.
We also performed simulations at constant force, which corresponds
to the limit of infinitely weak springs.
However, the unwinding of the proteins occurs in an "all or nothing"
fashion in this limit, and little information can be extracted.
\\

There are many ways to pick the pulling direction,
and all of our mechanical results correspond to a case in which the
extension is implemented parallel to the initial end-to-end vector.
In most cases, we pull the spring very slowly -- at a constant rate of
$v_p\;= \; 0.005 $\AA$/\tau$.
There is actually very little dependence of the results
on pulling rate until one considers rapid rates.
For instance, increasing $v_p$ by a factor of 50 
produces almost no change in the force.
Substantial rate dependence begins when $v_p$ is increased
by a factor of 100 to 0.5 \AA$/\tau$, and this 
case is denoted as a "fast" stretch in the following section.
The instantaneous pulling force $F$ is the extension of the pulling
spring times the spring constant $k$.
Plotted values of $F$ are averaged over 1$\tau$.
The standard pulling velocity is low enough that the force
equilibrates along the chain and almost the same force is obtained
from the extension of the spring whose end is fixed.
Drag terms lead to a significant
difference in these forces at higher velocities.
The force is plotted versus the cantilever
displacement $d\; =\; v_p t$, where $t$ is the total pulling time.\\


\section*{RESULTS AND DISCUSSION}

\noindent
\underline{\bf $\alpha$-helix}

Figure 2 illustrates the process of mechanical unfolding for H16.
It clearly shows that unfolding starts at both ends
and then proceeds to the center.
This is precisely the ordering of events during
thermal folding \cite{Hoang} and {\em not} the inverse of this ordering
as seen for the C-terminal hairpin of protein G \cite{Rokhsar}.
However, the underlying reasons for the observed ordering during folding
and unfolding are different.
Folding starts at the ends because they diffuse more rapidly
and are thus more likely to fall into a contact situation,
while unfolding starts at the ends because
there are fewer binding forces there.\\

The force vs. cantilever displacement
curves are shown in Figure 3 for slow (solid lines) and
fast (dashed lines) displacement rates.
The curves are truncated
when the helix is fully stretched, and any further displacement
results in rapid growth in $F$ followed by rupture
of the protein backbone.
The upper (lower) panels are for the stiff (soft) pulling springs.
In both cases unfolding produces a sequence of stick-slip events.
The force rises linearly while the protein is trapped in a given
local energy minimum, and then drops rapidly as one or more contacts
breaks.
The slope of the upward rise is the
combined stiffness $k_{tot}$ of the protein $k_p$ and cantilever $k/2$.
Since the two are in series, $k_{tot}^{-1}=k_p^{-1}+2k^{-1}$.
In the soft spring case the cantilever dominates and the slope of the
upward ramps is $k/2$.
For the stiff cantilever case the internal stiffness of the protein
dominates.
Variations within and between local minima lead to changes in the
slope of the ramps, with $k_p$ varying between about 0.3 and 0.6
$\epsilon/$\AA$^2$.
Once $k$ is larger than these values it has little influence on the
curves.

Each upward ramp ends when one or more contacts break.
The force drops sharply until the protein reaches a new metastable
state and a new upward ramp begins.
In the low velocity case (solid lines), $v_p$ is much lower than
the velocities produced by contact breaking, and rupture occurs at
a nearly constant cantilever position.
In this limit, the force drop is roughly equal to $k_{tot}$ times the
change in protein length during the jump between
metastable states.
For a stiff cantilever (top panel),
the failure of each contact produces a large drop in the force.
The first two peaks correspond to breaking of the two end contacts.
The force is lower than for later events
because the ends have fewer native contacts.
Rupturing of the next series of bonds proceeds in an essentially
periodic pattern because each ruptured bond has the same environment.
When the remaining helical segment is short enough, failure affects
bonds across its entire length, leading to two higher peaks.
In the final stages (i.e. $d > 28$\AA\ ),
all the coils have been broken and the
series of small force peaks
is due to breaking of higher order bonds.

When a soft spring is used, the drop in force due to each event
is smaller.
If the threshold force for an event is lower than that for the previous
event, the force may not drop below this threshold.
This can cause several bond ruptures to accumulate
into a single orchestrated event.
The low velocity curve in the lower panel of Fig. 3 has the same
initial sequence of peaks as the upper panel:
Two small peaks are followed by several at the same higher force.
However, those later peaks that are well below preceding peaks in the top
panel are absent in the bottom panel.
The strength of the contacts broken in these stages would be difficult
to extract if a soft cantilever were used.\\

When the pulling velocity is comparable to the rapid motions produced
by bond rupture, the cantilever motion can produce a substantial change
in force during an unfolding event.
This can also cause events to accumulate as shown in both panels.
The increase in speed also produces a larger drag force from the
surrounding solution (represented by the Langevin damping).
This shifts the force curves to higher values.\\

The optimal temperature for folding 
of H16 has been established to be
$T_{min}=0.3\epsilon / k_B$ \cite{Hoang}. The sequencing of thermal folding
and unfolding events
at $T_{min}$ is shown in Figures 4 and 5. The former Figure considers
establishment of the contacts of the $i,i+4$ type, i.e. the
hydrogen-bonded contacts, whereas the latter is for the $i,i+3$
contacts. The time for establishing of a given contact is denoted by
$t_c$. These times are symmetrically arranged around the
center of the helix and are shortest at the ends.
We have also determined times for thermal unfolding, $t_u$, defined as 
times at which the contact is gone for the first time. Values of $t_u$ in
Figures 4 and 5 are
averaged over 1500 different trajectories which all start in the native 
state. We notice that
$t_u$ is an order of magnitude longer than $t_c$ but it is also arranged
symmetrically around the center of the helix.

Figures 4 and 5 also show the displacements, $d_u$, at which each bond
ruptures during mechanical unfolding for stiff (closed circles)
and soft (open circles) cantilevers.
These curves do not have the same symmetry as the thermal folding
and unfolding curves.
As noted above, the end bonds break first because they have fewer
native contacts.
Subsequent bonds have the same number of contacts and should break
at the same force.
However,
the bonds near the pulling end (large $i$) tend to break
first due to the presence of a small extra drag force.
This is independent of the nature of the
cantilever except that the soft spring yields uniformly larger
$d_u$ at which a bond breaks.\\

Despite the lack of symmetry in the mechanical data, the folding times and
contact breaking distances are clearly correlated. This is shown 
in Figures 6 and 7 for the stiff and soft spring respectively. In each
figure, contacts breaking at later times tend to break at larger
displacements.\\

\noindent
\underline{\bf Two helices in tandem}

We now consider two H16 helices connected in series by one extra
peptide bond and stretched from one end.
Figure 8 shows a snapshot of a partially unfolded
tandem conformation. It indicates that the two helices unfold simultaneously
with some phase shift between them. This is also seen in the $F - d$ curves
shown in Figure 9 where the stick-slip patterns essentially double each feature
seen in Figure 3.
This behavior is quite distinct from what happens
when stretching titin where the domains unfold one at a time \cite{second}.
The basic reason that the helices unfold simultaneously is that the force
to break contacts rises smoothly during the unfolding process.
The heights of the force peaks
only drop in the very late stages of growth when the
coils are all gone.
In the case of titin,
one of the early peaks is higher than subsequent peaks.
Once this contact breaks in one of the repeat units, there is a series of
weaker bonds that can continue to rupture within that unit.
These contact failures
keep the force from rising back to a level that would initiate
failure of the strong bonds in other repeat units.\\

The simultaneous unwinding of the two helices is also seen in Figure 10
which is an analog of Figure 4 for the single helix
(minus the data on thermal unfolding).
The distance for contact rupture (for $i,i+4$ contacts) through 
stretching shows two skewed peaks, each centered in the
vicinity of the centers of the individual helices.
In contrast, the average times for unfolding at $T=0.3 \epsilon/k_B$
are peaked not at the centers of the helices but at 
the very center of the whole system, i.e. around the peptide bond
that connects the helices. Thus the simple correlation between
$d_u$ and $t_c$ that was seen in Figures 6 and 7 is lost.
Instead one finds a two-legged correlation
that is shown in Figure 11.
Note also that all of the contacts (all are short
ranged and are grouped into three types: $i,i+2$, $i,i+3$, and $i,i+4$)
break throughout the full range of the displacement of the cantilever.
Some bonds of a given kind break early, some break late. We shall see
in the companion paper \cite{second} that failure of long range bonds shows
a definite correlation with the displacement.\\

\noindent
\underline{\bf $\beta$-hairpin}

The stretching of the $\beta$-hairpin B16, shown in Figure 12, consists
of a gradual removal of the "rungs" of the "ladder" that 
form the hairpin, 
starting from the free ends.
Physically, these rungs represent hydrogen bonds
and they correspond to contacts 1-16, 2-15, 3-14, ..., 7-10.
There are other contact forces in our Go model and they provide
further stabilization of the structure. These other bonds bind bead
1 with bead 15, bead 2 with 14 and 16, etc.\\

Plots of $F$ versus $d$ during unfolding at low $v_p$
are shown in Figure 13. All show regular stick-slip features.
In this respect, our results are very similar to those
obtained by Bryant et al. \cite{Rokhsar} with full atom simulations.
Thus our simplified model reproduces the features present in the
more realistic calculation. Furthermore, since our model incorporates
the native conformation but not the hydrophobic or polar properties
of the amino acids, we suggest that the latter are not explicitly
crucial in the mechanical unfolding of the hairpin.
The stiff and soft springs produce the same sequence of stick-slip
peaks, but the slope of the ramps and depth of the drops are smaller
for the soft spring.
After the first peak, peaks come in pairs where the second peak
has a lower height.
When a very soft spring is used, these pairs merge into single large
events as described above.
\\ 

The folding properties of B16 are illustrated in Figure 14.
This system has been studied in detail in ref. \cite{Hoang}, where the
native basin has been accurately determined through a "shape distortion
technique" \cite{Li} which produces $T_{min}$ of order 0.07$\epsilon /k_B$.
If the folding criterion is based on just establishing the native
contacts, then, in the case of B16, there is a very broad dependence
of the folding time on temperature and the kinetics 
of folding at 0.07$\epsilon /k_B$
is almost the same as at, say, 0.3 $\epsilon /k_B$.
Nevertheless we study the system at the previously determined $T_{min}$.
Note that even with the contact-based criterion for folding, the folding time
for B16 is still considerably longer than for H16.\\

Figure 14 shows that the sequencing of folding events in B16 is exactly
opposite to the succession of contact breakage upon stretching:
B16 starts folding from the turn
(the result that has been found both experimentally \cite{Munoz}
and theoretically \cite{Hoang,Klimov})
whereas both mechanical and thermal
unfolding start at the free ends. Thus, in contrast to the
$\alpha $-helix, the mechanical unfolding of the $\beta$-hairpin is the
inverse of the folding process.\\

Figure 15 shows $d_u$
as a function of the time needed to establish the contact during folding.
Here, in addition to the "rung" contacts, the remaining contacts are also
shown. Since contacts rupture at a fixed force, the
soft spring data are shifted to larger displacements
than the the stiff spring data.
However both sets of data show a clear
anticorrelation between thermal folding and mechanical
unfolding that is in a sharp contrast to the results for
the $\alpha$-helix.\\

\noindent
\underline{\bf Irreversibility length}

We now consider pulling of a protein at a constant slow rate and then
releasing it. We ask what is the time required to fold back
to the native state at $T=0$. There must be a limit to the extension
beyond which the protein misfolds on release.
Figure 16 shows that this limit indeed exists.
The dependence on cantilever stiffness is minimized by plotting the refolding
times against the end-to-end distance $L$
of the protein rather than the cantilever
displacement.
For both stiffnesses the refolding times are found to be
non-monotonic functions of $L$.
We interpret this as being due to inertial effects.
The more stretched the protein is with a given set of contacts, the
more potential energy is available.
When the protein is released, the energy is converted into kinetic
energy that speeds the contraction of the protein and aids it in getting
over subsequent energy barriers.\\

We identify the irreversibility length $L_{ir}$ with the maximum
value of $L$ where refolding occurs.
For H16, $L_{ir}$ is about 37\AA , or 
1.6 times the native state end-to-end distance of 22.62 \AA.
The change in length is 14.4\AA\ which is very close to the displacement
of the stiff cantilever at the onset of irreversibility $d_{ir}=14.9$\AA\ .
The displacement of the soft cantilever, $d_{ir}=37.4$\AA\ ,
is larger because the cantilever stretches more in order to apply
enough force to reach $L_{ir}$.
Examining Figure 3, we see that both values of $d_{ir}$ correspond
to the displacement after the sixth peak in the respective force curve.
Thus the same set of broken bonds is required to produce irreversibility
for either cantilever stiffness.
\\

For B16, the native $L$ is only 5 \AA\ 
and the stretching factor to $L_{ir}$ is substantially larger, around 11.6.
The values of $d_{ir}$ for stiff and soft cantilevers are
$d_{ir}=52.9$ and 65.9\AA, respectively.
From Figure 13 we see that in both cases the irreversibility point is just
past the last peak in the force curve.
Since the protein is fully stretched at this point,
any native contacts are enough to ensure refolding.
\\

The misfolded conformations, that are obtained on refolding beyond the
threshold, are shown in Figure 17. In the case of B16, the turn
region freezes into the wrong configuration which is almost straight.
In the case of H16, the first turn coils
with the wrong chirality.\\

\section*{CONCLUSIONS}

We have studied the force - displacement curves
for secondary structures of proteins for two models of 
cantilever stiffness and several pulling speeds.
A series of stick-slip events is observed as contacts break.
Stiff cantilevers pulled at low rates provide the most detailed
information about the breaking of individual contacts.
Multiple contact ruptures merge into single events when the
stiffness is decreased or the speed is increased.\\

The simple expectation that mechanical unraveling should proceed
in the inverse order from thermal folding
is only confirmed in the case of the $\beta$-hairpin.
In the case of the $\alpha$-helix, unraveling and folding follow
the same order.
When multiple helices are connected in tandem, the correlation
becomes even more complex.
The two helices unravel simultaneously with each
helix uncoiling from both of its ends.
In contrast, folding occurs first at the outer ends of the pair of helices.
In general there is no reason to expect
a simple correlation between thermal folding and
mechanical unfolding of proteins.
In the following paper we examine
similar issues of mechanical-thermal correlations for a protein
with a significant number of long-ranged contacts.

\section*{ACKNOWLEDGMENTS}
We appreciate discussions with J. R. Banavar and T. Woolf 
which helped in motivating this research. This work was funded
by National Science Foundation Grant DMR-0083286, the Theoretical
Interdisciplinary Physics and Astrophysics Center at Johns Hopkins
University, and KBN.




\begin{figure}
\epsfxsize=2.5in
\vspace*{2.5cm}
\centerline{\epsffile{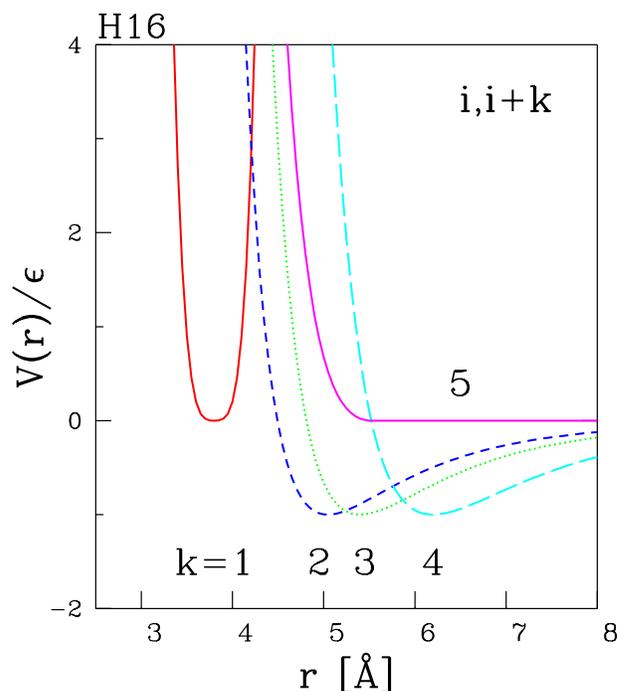}}
\vspace*{2cm}
\caption{ 
The potentials used to construct a Go model of the $\alpha$-helix H16.
The interactions are between the beads $i$ and $i+k$.
For $k$=1 this is the anharmonic tethering potential. The contact
corresponding to $k$=5 is non-native and is thus purely repulsive.
The remaining contact interactions are of the Lennard-Jones form.
}
\end{figure}
\newpage

\begin{figure}
\vspace*{-2cm}
\epsfxsize=2.5in
\centerline{\epsffile{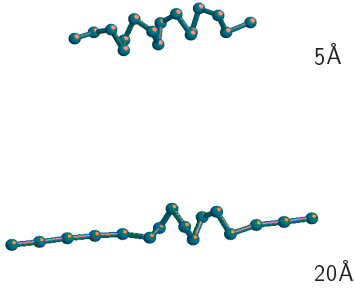}}
\vspace*{1cm}
\caption{
Snapshot pictures  of stretching of the $\alpha$-helix 
H16. The left end is anchored elastically and the right 
end is pulled by the stiff cantilever. The numbers indicate the
cantilever displacements. In the upper figure the helix is still almost fully
folded.
 }
\end{figure}

\begin{figure}
\epsfxsize=2.5in
\vspace*{0.5cm}
\centerline{\epsffile{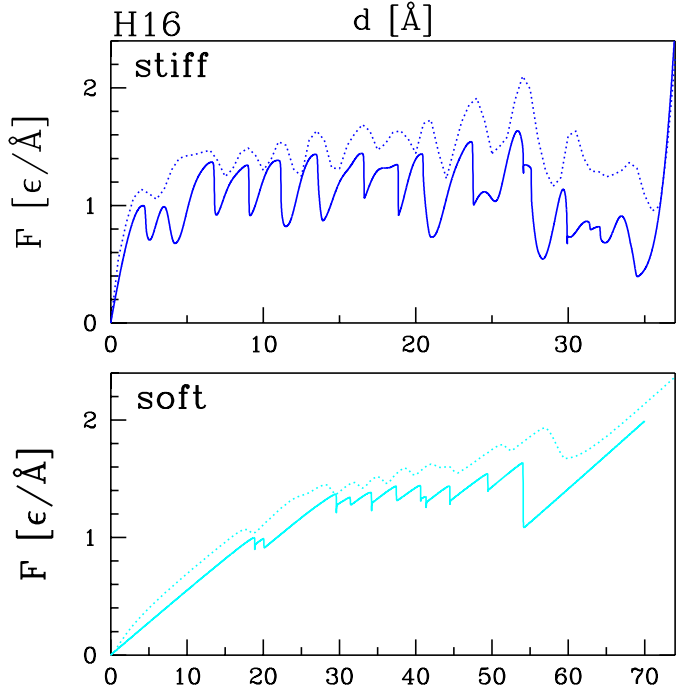}}
\vspace*{2.5cm}
\caption{
Force vs. displacement for H16 with stiff and soft cantilevers.
The solid (dotted) lines correspond to the slow (fast) pulling rates.
}
\end{figure}

\begin{figure}
\vspace*{3cm}
\epsfxsize=2.5in
\centerline{\epsffile{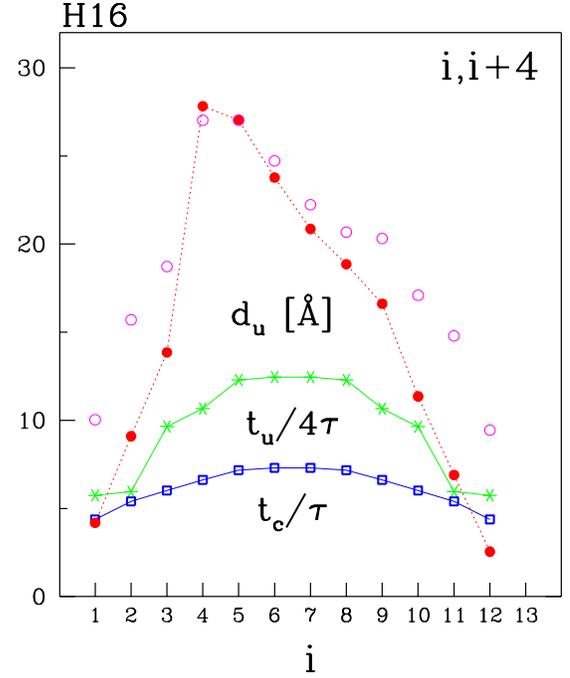}}
\vspace*{3cm}
\caption{
Sequencing of events as measured through forming or breaking bonds of the
$i,i+4$ kind. The $x$-axis shows $i$ -- the monomer number
along the chain. The moving cantilever is attached to $i$=16. 
The $y$-axis shows $t_c$, $t_u$, and $d_u$.
The first quantity is the mean time needed to establish the
contact at $T_{min}$ (based on 1000 trajectories). The second quantity is
the first unfolding time at $T_{min}$ for a bond under the conditions of
no pulling force (based on 1500 trajectories).
The third quantity, denoted by the circles,
is the displacement where the contact is broken during mechanical unfolding
at $T$=0.
The solid (open) circles correspond to slow pulling by the 
stiff (soft) spring. Values of $d_u$ for the soft cantilever
are divided by 2.
Overall, the size of the symbol is a measure
of the error bars, and all lines are guides to the eye.
}
\end{figure}
\newpage

\begin{figure}
\epsfxsize=2.5in
\centerline{\epsffile{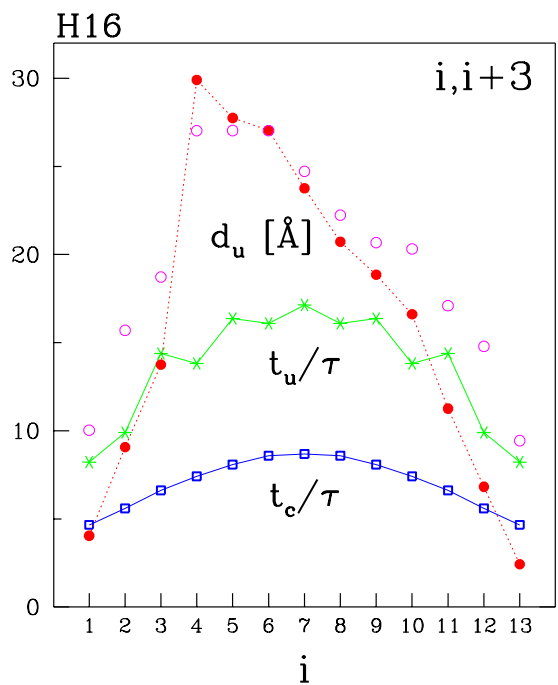}}
\vspace*{2cm}
\caption{
Same as in Figure 4 but for the bonds of the $i,i+3$ kind.
Values of $d_u$ for the soft cantilever are divided by 2.
}
\end{figure}

\begin{figure}
\vspace*{0.5cm}
\epsfxsize=2.5in
\centerline{\epsffile{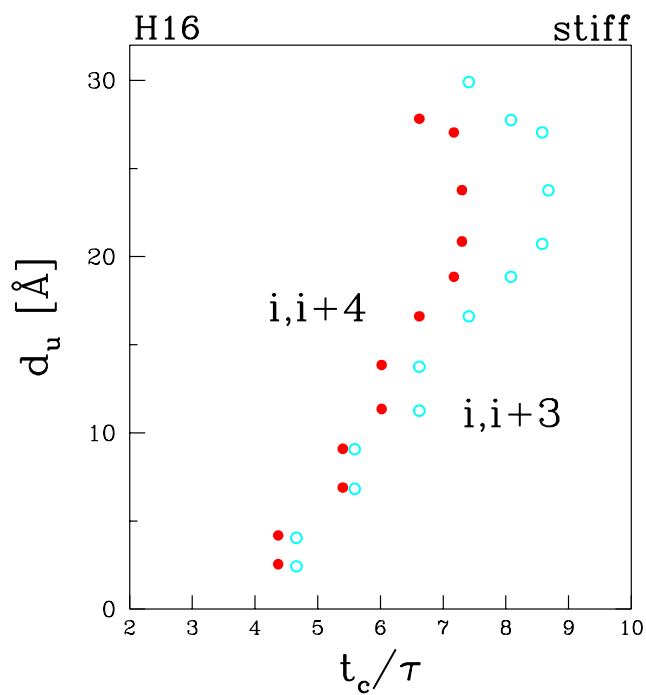}}
\vspace*{2cm}
\caption{
Stretching distances at which a bond rupture takes place (from Figures
4 and 5) plotted vs. average time needed to establish contact on folding.
This is the case of a stiff spring which is being pulled slowly.
}
\end{figure}

\begin{figure}
\epsfxsize=2.5in
\centerline{\epsffile{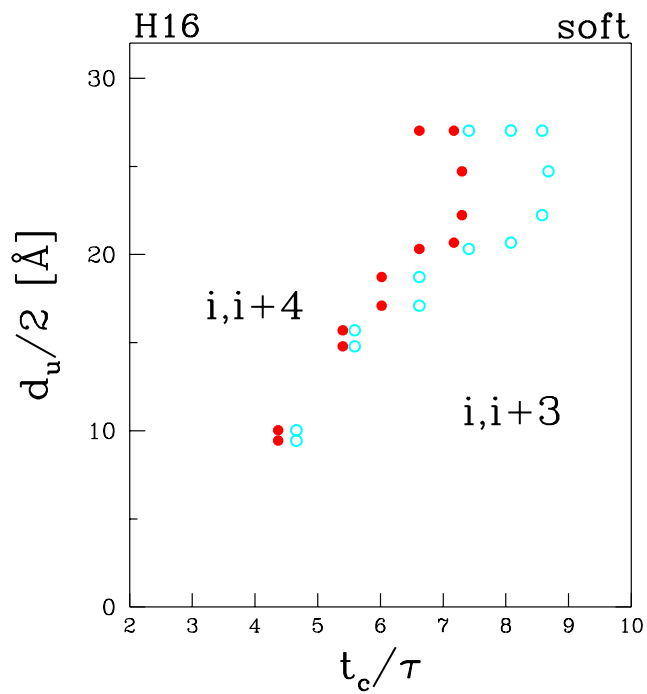}}
\vspace*{2cm}
\caption{
Same as in Figure 6 but for the soft pulling spring.
}
\end{figure}

\begin{figure}
\epsfxsize=2.5in
\vspace*{0.5cm}
\centerline{\epsffile{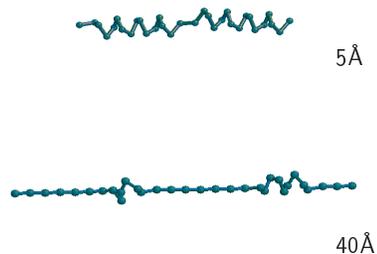}}
\caption{
Conformation of two $\alpha$-helices connected in series after
moving the cantilever by the distance indicated.
}
\end{figure}

\begin{figure}
\epsfxsize=2.5in
\centerline{\epsffile{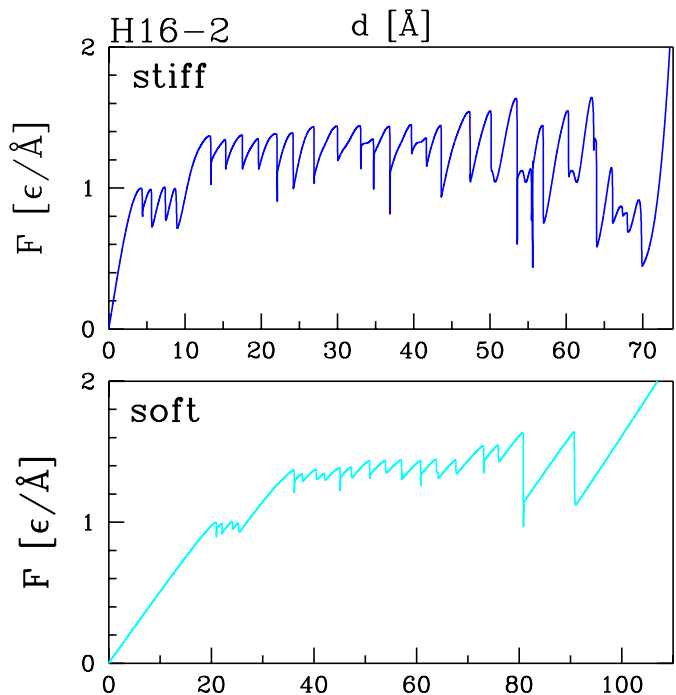}}
\vspace*{2cm}
\caption{
Force vs. displacement for H16-2 -- two $\alpha$-helices H16 
connected in series and pulled slowly.
}
\end{figure}

\begin{figure}
\epsfxsize=2.5in
\vspace*{1.5cm}
\centerline{\epsffile{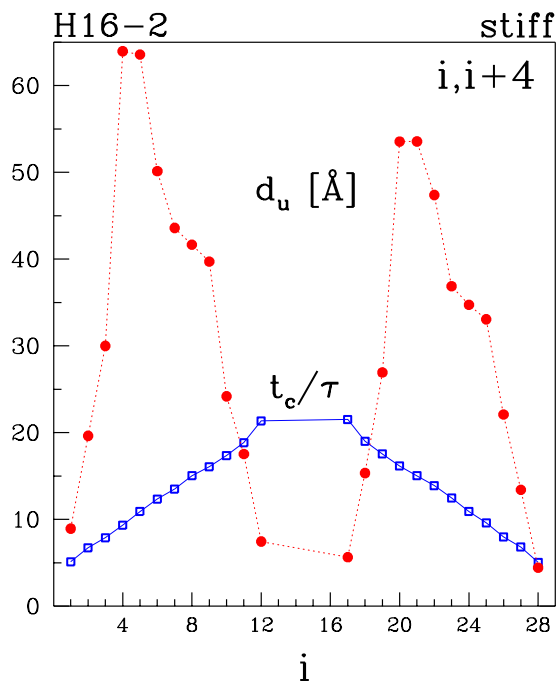}}
\vspace*{2cm}
\caption{
Sequencing of events in H16-2 as measured through the bonds of the
$i,i+4$ kind. The symbols have meanings as in Figure 4.
1000 trajectories were used in the studies of folding.
}
\end{figure}

\begin{figure}
\epsfxsize=2.5in
\centerline{\epsffile{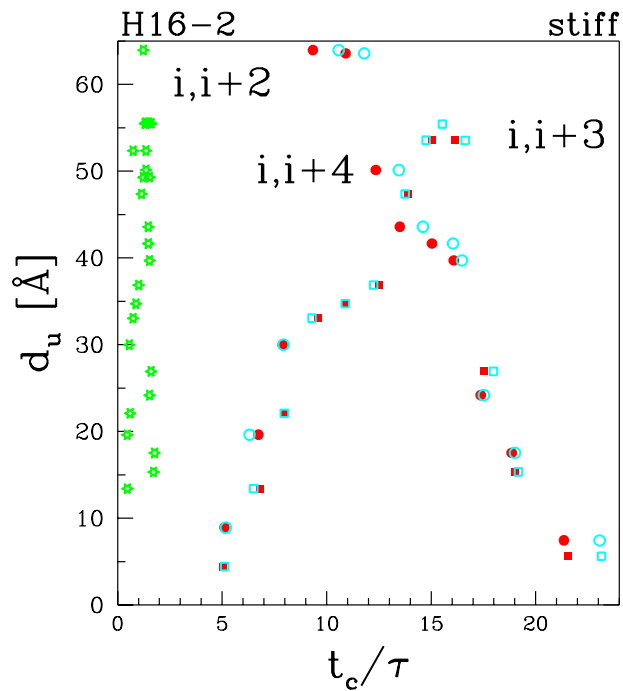}}
\vspace*{2cm}
\caption{
Stretching distances at which a bond rupture in H16-2 takes place
vs. average time needed to establish contact on folding.
The cantilever is stiff and pulled slowly.
The contacts corresponding to the $i,i+2$ type
are shown by stars. The remaining symbols differentiate between
the contacts present in the first helix (circles) and those present in the
second helix (squares) -- the one which is closer to the cantilever.
The open circles and squares correspond to the contacts of the $i,i+3$
type and the filled circles and squares to $i,i+4$.
}
\end{figure}

\begin{figure}
\epsfxsize=2.5in
\centerline{\epsffile{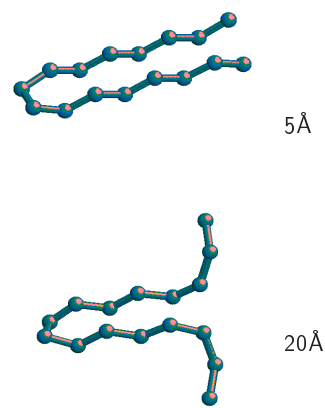}}
\caption{
Snapshot pictures of stretching of the $\beta$-hairpin B16 for $d$
equal to 5 and 20 \AA .
}
\end{figure}

\begin{figure}
\epsfxsize=2.5in
\centerline{\epsffile{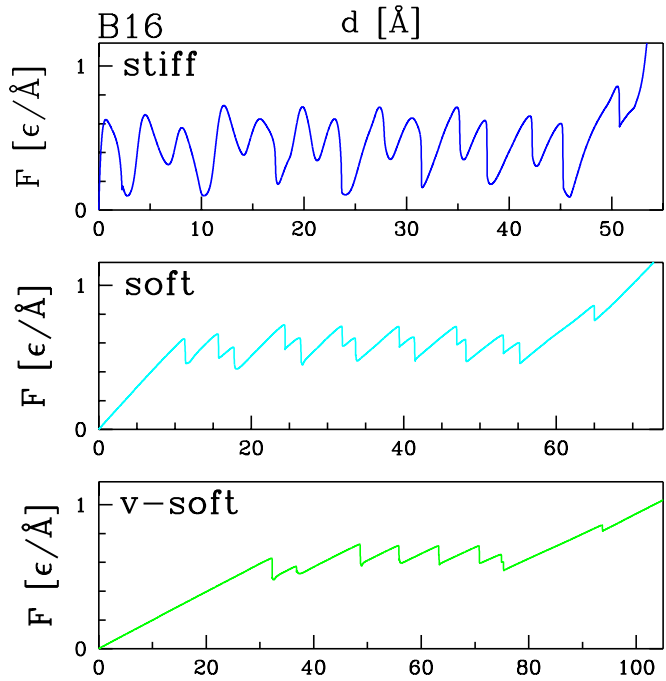}}
\vspace*{2cm}
\caption{
The force vs. displacement curves for the $\beta$-hairpin B16
obtained at slow pulling velocities.
}
\end{figure}

\begin{figure}
\epsfxsize=2.5in
\vspace*{0.5cm}
\centerline{\epsffile{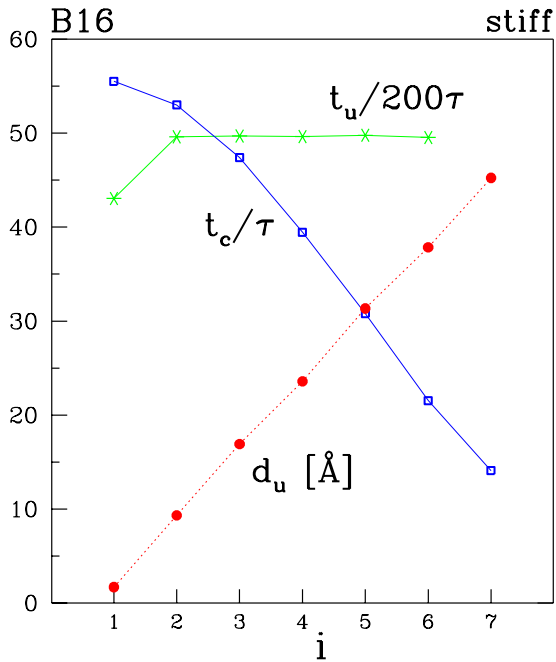}}
\vspace*{2cm}
\caption{
Similar to Figure 4 but for the hydrogen rung-like bonds in B16.
The bonds are identified by the index $i$ that they connect to.
The thermal data are based on 1000 trajectories and are collected
at $T_{min}= 0.07\epsilon /k_B$.
The flat character of the data corresponding to thermal unfolding
is expected to turn into a steeper dependence at higher temperatures.
}
\end{figure}

\begin{figure}
\epsfxsize=2.5in
\centerline{\epsffile{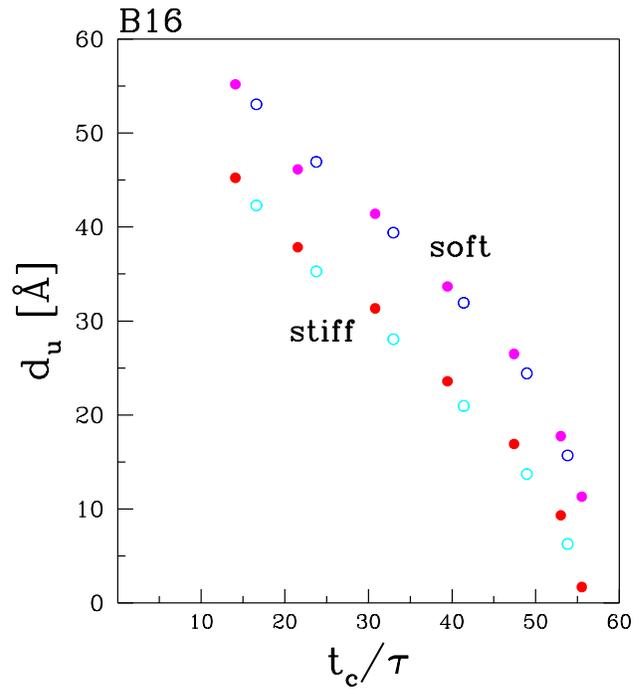}}
\vspace*{2cm}
\caption{
Stretching distances at which a bond rupture in B16 takes place
vs. average time needed to establish contact on folding.
The cantilever is pulled slowly. The solid circles correspond
to the "rung" contacts. The open circles correspond to the
remaining contacts. There are degeneracies related to these other contacts.
For instance, 2-14 forms and breaks at essentially the same moment
(statistically) as 3-15. Examples of other such pairs are
3-13 with 4-14 and 6-12 with 5-11. 
}
\end{figure}
\newpage

\begin{figure}
\epsfxsize=2.5in
\centerline{\epsffile{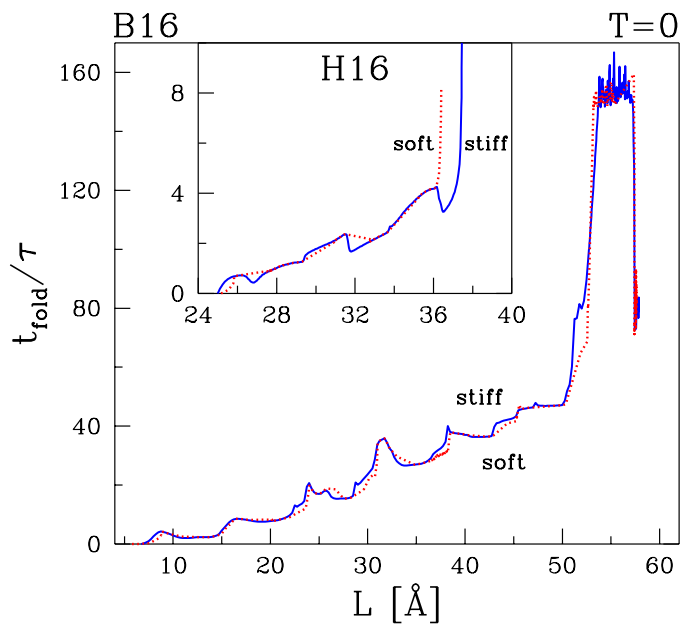}}
\vspace*{2cm}
\caption{
Refolding times after stretching to the indicated end-to-end distance.
The main figure is for B16 and the inset for H16. The solid (dotted) lines
correspond to pulling by a stiff (soft) cantilever. The curves
end at $L_{ir}$.
To the right of the data points shown,
the protein does not return to its native state.
For B16, the corresponding threshold values of the
tip displacement, $d_{ir}$,
are equal to 52.9$\AA$ and 65.9$\AA$ for the stiff and soft 
cases respectively. For H16, the values of $d_{ir}$
for the stiff and soft cantilevers are 14.9$\AA$ and 37.4$\AA$,
respectively.
}
\end{figure}

\begin{figure}
\epsfxsize=2.5in
\centerline{\epsffile{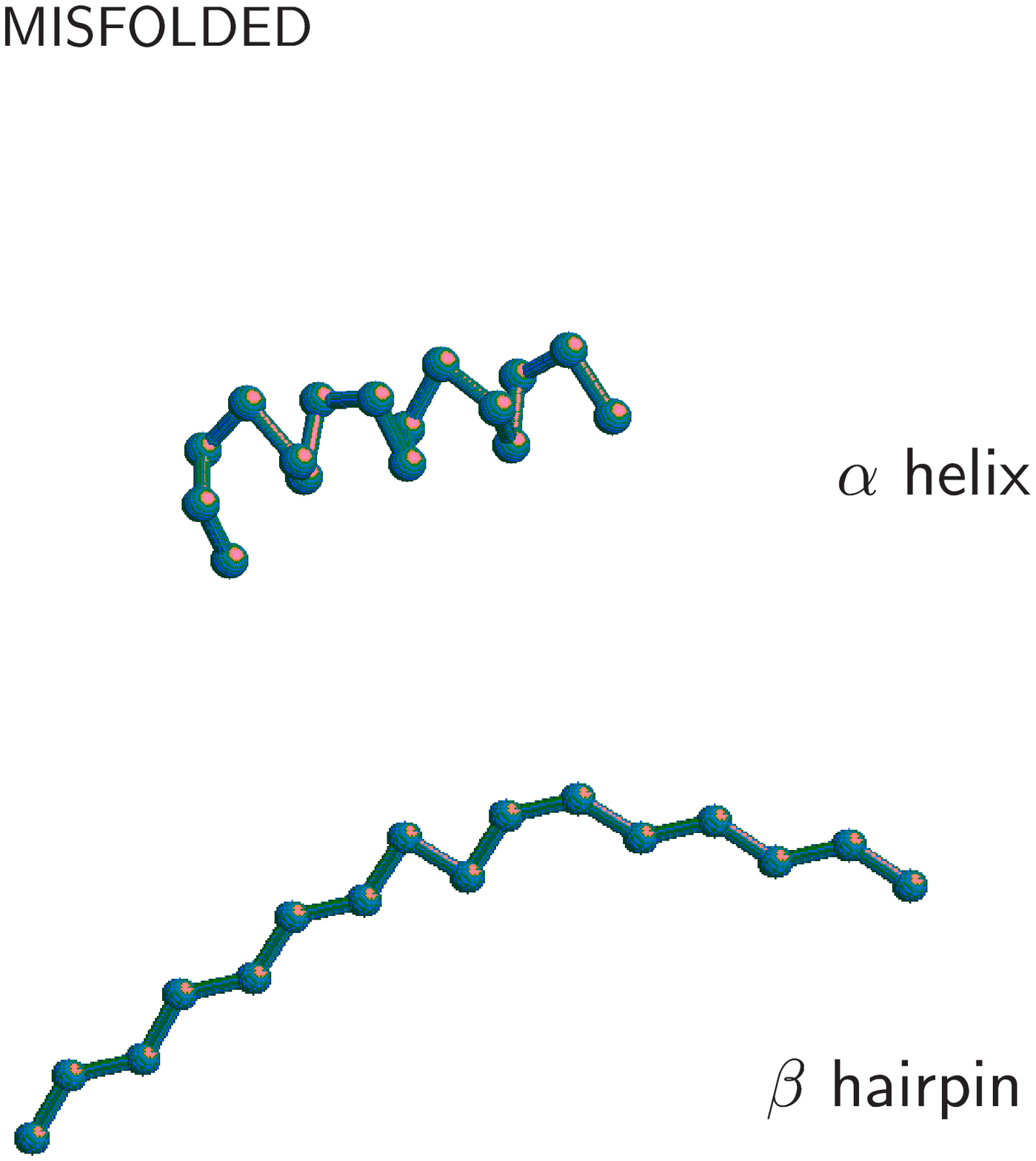}}
\caption{
Conformations corresponding to the misfolded proteins after stretching
just beyond the irreversibility threshold.
}
\end{figure}
\end{document}